# Metrics for popularity bias in dynamic recommender systems

Valentijn Braun[1][a], Debarati Bhaumik [1][b] and Diptish Dey[1][c]
[1]*Amsterdam University of Applied Sciences, the Netherlands*
*{valentijn.braun, d.bhaumik, d.dey2}@hva.nl*

Abstract: Albeit the widespread application of recommender systems (RecSys) in our daily lives, rather limited research has been done on quantifying unfairness and biases present in such systems. Prior work largely focuses on determining whether a RecSys is discriminating or not but does not compute the amount of bias present in these systems. Biased recommendations may lead to decisions that can potentially have adverse effects on individuals, sensitive user groups, and society. Hence, it is important to quantify these biases for fair and safe commercial applications of these systems. This paper focuses on quantifying popularity bias that stems directly from the output of RecSys models, leading to over recommendation of popular items that are likely to be misaligned with user preferences. Four metrics to quantify popularity bias in RescSys over time in dynamic setting across different sensitive user groups have been proposed. These metrics have been demonstrated for four collaborative filtering based RecSys algorithms trained on two commonly used benchmark datasets in the literature. Results obtained show that the metrics proposed provide a comprehensive understanding of growing disparities in treatment between sensitive groups over time when used conjointly.

## 1 INTRODUCTION

RecSys have become an integral part of our daily lives, influencing the products we buy, the movies we watch, the music we listen to, so on and so forth (Lu et al., 2015). These systems aim to predict users' preferences and provide personalized recommendations by analysing their past behaviour, preferences, and interactions (Lü et al., 2012). The explosion of e-commerce and the growth of online platforms resulted in RecSys becoming essential tools for businesses to increase user engagement and customer loyalty (Khanal et al., 2020). Similarly, RecSys are also finding its applications in sensitive sectors such as law enforcement (Oswald et al., 2018), health care (Schäfer et al., 2017), and human resources (Vogiatzis & Kyriakidou, 2021); usage in these sensitive sectors necessitates that recommendations provided can be explained, evaluated, and demonstrably unbiased & fair.

Contemporary research in RecSys has focussed on improving accuracy and processing speed (Chen et al., 2023). Meanwhile, RecSys algorithms continue to be trained on data with real-world user behaviour, which is shown to contain various biases such as representation bias and measurement bias (Mehrabi et al., 2021), despite that research on biases in RecSys lacks consensus on definition of bias (Chen et al., 2023; Deldjoo et al., 2023). Additionally, prior works on algorithmic fairness focus primarily on defining conditions for fairness to answer the question "is an algorithm unfair?", but do not provide evaluation metrics for unfairness to answer how unfair an algorithm is (Speicher et al. (2018)). This is further supported by Lin et al. (2022), emphasising on how to quantify bias in RecSys still remains understudied.

Bias, in context of RecSys, can be viewed as recommendations provided by such systems that may potentially lead to discrimination towards certain items, groups or individuals based on factors such as demographics, item popularity, personal preferences or historical data (Chen et al., 2021). The various types of biases that exist in RecSys can be categorized into four categories: data bias, model bias, results bias, and amplifying biases (Chen et al., 2023).

[a] https://orcid.org/0009-0009-5441-4528

[b] https://orcid.org/0000-0002-5457-6481

[c] https://orcid.org/0000-0003-3913-2185

Data bias refers to biases that are present in input data used to train RecSys algorithms. This consists of selection bias, conformity bias, exposure bias, and position bias. Selection bias arises from users' freedom to choose which items to rate, leading to users tending to select and rate items that they like and are more likely to rate particularly good or bad items (Marlin et al., 2007). Conformity bias involves users rating items in line with group behaviour and not to their true preferences (Liu et al., 2016). Exposure bias results from disproportionate presentation of unpopular items to users compared to popular items (Liu et al., 2020), whereas position bias occurs when item positions in a list of recommended items influence user interaction (Collins et al., 2018).

Model bias represents inductive biases that are purposefully added to the model design in order to achieve desirable results which cannot be derived from training data (Chen et al., 2023).

Results bias pertains to biases that originate directly from output of RecSys models. Such biased recommendations lead to: (i) popularity bias, where popular items are recommended with higher propensity, potentially mismatching user preferences, and (ii) unfairness, in which discriminatory recommendations are provided to certain individuals or groups with specific attributes like race or gender (Mehrabi et al., 2021; Ekstrand et al., 2018).

Amplifying biases occur when existing biases present in the data, model or results are amplified unintentionally, thus, intensifying disparities. This effect involves self-reinforcing feedback loops where recommendations reinforce existing preferences and perpetuate bias (Mansoury et al., 2020).

Biased recommendations have varied negative consequences (Kordzadeh & Ghasemaghaei, 2022); Popularity bias might undermine user's interactions with items that are unpopular and prevent them from becoming popular (Baeza-Yates, 2020). Mehrotra et al. (2018) illustrated that a small number of popular artists on Spotify get an overwhelmingly larger number of listens, resulting in an unfavourable consequence for the remaining less renowned musicians. Similarly, it has been studied that popularity and demographic biases led to users with different ages, genders, and/or demographics receiving recommendations with significant differences in accuracy (Ekstrand et al., 2018). Since, these factors can potentially lead to discrimination and unfairness towards individuals or groups, it is important to quantify popularity bias and unfairness in RecSys (Deldjoo et al., 2023; Ekstrand et al., 2018; Mehrabi et al., 2021).

Studies done on quantifying popularity bias mostly focus either on static settings or at a global level (Ahanger et al., 2022; Abdollahpouri et al., 2019; Ekstrand et al., 2018). However, in real-life applications, unfairness may only become apparent over time across different user groups. To measure unfairness, this paper proposes metrics to quantify popularity bias in dynamic settings across sensitive user groups in RecSys.

In section 2, metrics currently used to measure popularity bias and their limitations are discussed. In section 3, the proposed metrics for quantifying popularity bias over time across various sensitive user groups are presented. In section 4 the proposed metrics are demonstrated using two commonly used datasets in literature for two sensitive user groups, males and females. In section 5 and 6, conclusions and future work are discussed, respectively.

## 2 POPULARITY BIAS IN RECSYS

When training algorithms on long-tailed data, RecSys models tend to give higher scores to items that are more popular, resulting in popular items being recommended with higher propensity than their actual popularity (Abdollahpouri & Mansoury, 2020). This results in recommendations provide by a RecSys based on a biased selection of items that do not align with the user's actual preferences, thereby, negatively impacting the user experience (Bhadani, 2021). Additionally, if popularity bias is ignored, a negative feedback loop can result in popular items getting even more popular (Zhu et al., 2021).

It is important to note that popularity bias is not always harmful. Item popularity is not only a result of conformity, where people tend to behave similarly to others within a group but can also result from the item being of high quality. This implies that leveraging popularity bias appropriately into a RecSys may improve its performance (Zhao et al., 2022).

### 2.1 Static versus Dynamic Setting

Studies conducted on evaluating fairness in RecSys use either a static or dynamic setting. Whereas static refers to data usage at a single point of time, dynamic setting includes usage of user interaction data over time including feedback interactions; the latter being closer to real-life implementations. Access to real-life dynamic data is a challenge. As a result, approximately 85% percent of recent studies on

RecSys are performed on static data (Deldjoo et al., 2023). Evaluating biases in RecSys within static settings may lead to under-representation of unfairness as it may only surface over time.

To this end, to measure biases in RecSys without access to real dynamic data from live platforms, user interactions must be simulated from static datasets. A common approach to transform a static dataset into a dynamic one by simulating new (dynamic) interactions (Aridor et al., 2020; Chong & Abeliuk, 2019; Zhu et al., 2021; Khenissi et al., 2020) is deployed in this paper. This approach uses the assumption that users interact with their top-N recommendations and appends these interactions to the static dataset for a predefined number of iterations. An iteration is a step in which a RecSys algorithm is trained to provide top-N recommendations to each user and their interaction is simulated (see section 4.2).

## 2.2 Individual versus Group Fairness

Concepts in algorithmic fairness can be categorized into two groups: individual- and group- fairness (Dwork et al., 2012). Individual fairness refers to the principle that similar individuals should receive similar predictions or outcomes from a machine learning model, ensuring that decisions are consistent across comparable cases (Zemel et al., 2013). On the other hand, group fairness focuses on preventing unfair discrimination against specific demographic or social groups, aiming to ensure equitable outcomes at a larger societal level (Luong et al., 2011).

Fairness metrics that contain a subgroup decomposability property (e.g., generalised entropy) can be used to decompose the overall individual-level unfairness into two components, namely between-group and within-group (un)fairness (Shorrocks, 1984). It has been observed that minimizing the level of between-group unfairness may, in fact, increase the level of within-group unfairness, leading to an increase in overall unfairness (Speicher et al., 2018). Recognizing the importance of considering both between-group and within-group (un)fairness, this paper measures both.

## 2.3 Current Metrics of Popularity Bias

Metrics that have been proposed in literature to measure popularity bias in RecSys are defined at a global level such as Gini coefficient (Deldjoo et al., 2023) or at group levels such as generalized entropy index (GEI) (Speicher et al., 2018) or in static settings such as delta group average popularity (ΔGAP) (Abdollahpouri et al., 2019). These metrics are summarized in this section.

### 2.3.1 Gini Coefficient

Gini coefficient, originally developed to serve as an indicator of income inequality within a society (Gini, 1936), has in recent years been applied to measuring popularity bias in RecSys (Abdollahpouri et al., 2021; Analytis et al., 2020; Chong & Abeliuk, 2019; Leonhardt et al., 2018; Lin et al., 2022; Sun et al., 2019; Zhu et al., 2021). Gini coefficient is used on the distribution of popularity score of each item in a dataset. Popularity score ($\phi_i$) of item $i$ is the ratio of the number of users that have interacted with an item by the total number of users, i.e.,

$$\phi_i = \frac{N_i}{N_U}, \qquad (1)$$

where $i$ is an item, $N_i$ is the number of users that interacted with the item, and $N_U$ the total number of users in the dataset (Abdollahpouri et al., 2019; Kowald & Lacic, 2022). By creating a distribution of item popularity scores from equation (1), Gini coefficient ($G$) is computed to quantify inequality within that distribution and is given by (Sun et al., 2019):

$$G = \frac{\sum_{i=1}^{n}(2i - n - 1)\phi_i}{n \sum_{i=1}^{n} \phi_i}, \qquad (2)$$

where $\phi_i$ is the popularity score of the $i^{th}$ item, where the rank of $\phi_i$ is taken in ascending order ($\phi_i \leq \phi_{i+1}$) and $n$ the number of items. Gini coefficients take values between 0 and 1, with 0 representing perfect equality and 1 representing maximum inequality.

The Gini coefficient as a fairness metric is used in both static (Abdollahpouri et al., 2021; Analytis et al., 2020; Leonhardt et al., 2018; Lin et al., 2022) and dynamic (Chong & Abeliuk, 2019; Sun et al., 2019; Zhu et al., 2021) settings. In a dynamic setting, increasing values usually indicate that certain items are being recommended more frequently than others (Deldjoo et al., 2023), suggesting a concentration of recommendations on a selection of items. Decreasing values suggest that a diverse range of items are being recommended to users.

However, Gini coefficient has only been used as an indicator of fairness at a global level in literature. As indicated in section 2.2, it is also important to assess the trade-off that exists amongst between-group and within-group (un)fairness. Hence, metrics to measure popularity bias using Gini coefficient for different sensitive user groups are proposed and demonstrated in sections 3.1 and 4.3.1 respectively.

### 2.3.2 Delta Group Average Popularity ($\Delta GAP$)

$\Delta GAP$, originally proposed by Abdollahpouri et al. (2019), is a metric that is used to measure popularity bias at a user group level by evaluating the interests of user groups towards popular items (Kowald et al., 2020; Yalcin & Bilge, 2021). It is based on the notion of calibration fairness, which assumes that fair recommendations should not deviate from historical data of users (Steck, 2018). Consequently, the objective is to minimise the difference between the recommendations and the profiles of users within a group, in which a user profile consists of all observed item-rating interactions of the user.

In general, $\Delta GAP$ computes the difference between the average popularity of items in group-recommendations to the average popularity of items in group-profiles (Wundervald, 2021). Based on the definition of item popularity as defined in equation (1), the group average popularity per user group $g$, $GAP(g)$, is defined as (Abdollahpouri et al., 2019):

$$GAP(g) := \frac{\sum_{u \in g} \frac{\sum_{i \in p_u} \phi_i}{|p_u|}}{|g|}, \quad (3)$$

where $g$ is a user group, $|g|$ the number of users in that group, $p_u$ is the list of items in the profile of a user $u$, $|p_u|$ is the number of items in the profile of user $u$, and $\phi_i$ is the popularity score of item $i$. In other words, $GAP(g)$ is the average of the average item popularity within each user profile belonging to a user group $g$.

To evaluate the difference between the recommendations and historical data of a specific user group, equation (3) is used to provide values for user profiles $GAP_p$ and for their corresponding recommendations $GAP_r$. $GAP_r$ is computed by changing $p_u$, the lists of observed interactions, with the lists of recommended items to users within that group. In an ideal situation of calibration fairness, the average popularity of the recommendations is equal to the average popularity of user profiles, i.e., $GAP_p = GAP_r$. Subsequently, Abdollahpouri et al. (2019) proposed $\Delta GAP$ to calculate the level of undesired popularity in group recommendations:

$$\Delta GAP(g) := \frac{GAP(g)_r - GAP(g)_p}{GAP(g)_p}. \quad (4)$$

The values of $\Delta GAP$ range from $-1$ to $\infty$ and can be interpreted as the relative difference of the average item popularity between user profiles and recommendations within a user group $g$. In this context, complete fairness is achieved when $\Delta GAP = 0$. As $GAP(g)_r$ tends to $0$, indicating that all recommended items are unpopular, $\Delta GAP$ tends to $-1$. Whereas, when $GAP(g)_p$ tends to 0, indicating that all items interacted with by users in a user group are unpopular, $\Delta GAP$ tends to $\infty$.

The current adaptations of $\Delta GAP$ to measure popularity bias in literature is only limited to static settings. Hence, a new metric $dynamic - \Delta GAP$, pertaining to more real-life dynamic settings is proposed in this paper (see section 3.2). Additionally, to measure between-group unfairness another metric in dynamic setting, $BetweenGroup - GAP$ is proposed in section 3.3.

### 2.3.3 Generalised Entropy Index ($GEI$)

$GEI$, a measure like the Gini coefficient, is used to quantify the degree of inequality or diversity within a distribution (Mussard et al., 2003). In the context of popularity bias, $GEI$ is used to measure inequality in the distribution of item popularity score. In contrast to the Gini coefficient, $GEI$ possesses the property of additive decomposability. For any division of a population into a set of non-overlapping groups, the $GEI$ over the entire population can be decomposed as the sum of a component for between-group unfairness and a component for within-group unfairness (Speicher et al., 2018). This allows to quantify how unfair an algorithm is towards sensitive groups within a population and visualise the trade-offs between individual-level and group-level fairness when debiasing RecSys models (Speicher et al., 2018).

## 3 PROPOSED METRICS FOR POPULARITY BIAS

Gini coefficient and $\Delta GAP$ are deployed in global contexts and in static settings respectively. $GEI$ has been used to measure popularity bias both at group levels and in dynamic settings, although Gini coefficient and $\Delta GAP$ remain more accessible measures due to their simpler structure (Wang et al., 2023). To measure time evolution of popularity bias and its differential treatment among sensitive groups, variants of Gini coefficient and $\Delta GAP$ are proposed in this paper, namely *Within-group-Gini coefficient*, *Dynamic-$\Delta GAP$*, and *Between-group GAP*. Additionally, another metric, *group-cosine similarity,* to quantify differential treatments between groups with similar characteristics has also been proposed.

### 3.1 Within-group-Gini coefficient

In RecSys, Gini coefficient is used as a measure of global inequality in the distribution of item popularities over time. We extend this application of Gini coefficient to dynamic settings to calculate *Within-group-Gini coefficient* and compare how it varies between different (sensitive) groups over time. *Within-group-Gini coefficient* is calculated as:

[1] The original dataset containing all user-item interactions is split into distinct datasets per group under consideration and encompassing all interactions concerning each group. Example of such groups are males and females.
[2] For each of the datasets pertaining to the groups, item popularity score (see equation (1)) is computed based on the respective interactions, thus generating separate distributions of item popularities for each group. For example, if we consider males and females to be the groups to assess, two separate distributions of item popularity score for males and females are generated.
[3] *Within-group-Gini coefficients* is calculated for each of the generated distributions of item popularities per group, using equation (2).
[4] Steps [1]-[3] are repeated per iteration, in which the top-N recommendations of all users are appended to the dataset to compute *Within-group-Gini coefficients* over time, pertaining to the dynamic setting adopted from Sun et al. (2019) and Zhu et al. (2021). For more details on this approach see Section 4.2.

By analysing the trends of *Within-group-Gini coefficients* of different user groups, it can be assessed whether the RecSys model exhibits differential treatment in recommendations between sensitive groups (such as males-vs-females) over time. This helps to understand if a RecSys model is offering less diverse recommendations to a particular group.

### 3.2 Dynamic-$\Delta GAP$

When reviewing the original proposal of $\Delta GAP$ by Abdollahpouri et al. (2019), it is expected that the top-N recommendations provided to a user in the testing dataset, contains only those items that a user has not previously interacted with. This approach has also been adopted by Kowald et al., (2020) for computing $\Delta GAP$. However, upon analysing their code base[4], it is found that in their approach, users are recommended items from the testing data with which they have already interacted. This leads to the following concerns:
- Recall that $\Delta GAP$ is defined such that a model is fair when recommendations align with historically observed data with the recommendations representing the performance of the model (see equation (4)). Therefore, providing recommendations that users have already interacted with, distorts the representation of $\Delta GAP$.
- For cases in which a user needs to be provided with more recommendations than the number of observed interactions in the testing data, the model returns all available interactions in the testing data for that user regardless of her predicted rating. This implies that when computing $\Delta GAP$, the user profile is compared to only a sample of items pertaining to that profile. This concern would primarily affect users in a small profile size, as there is a higher chance of having insufficient interactions in the testing dataset to generate an adequate number of predictions.

Setting aside these concerns, if recommendations are based on unobserved interactions, $\Delta GAP$ possesses the true potential to provide the intended insights on the level of popularity bias present in a RecSys model. Therefore, this paper proposes computing $\Delta GAP$ on unobserved interactions.

To compute *dynamic-$\Delta GAP$*, $\Delta GAP$ is computed using *unobserved interactions* over-time using simulated dynamic data with the approach described in section 2.1. Following are the steps for computing *dynamic-$\Delta GAP$*:
[1] Split original dataset into training and testing data.
[2] Define necessary user groups.
[3] Compute $GAP(g)_p$ (see Equation (3)) for each user group based on the training data.
[4] Train the specific RecSys algorithm on the training data.

---
[4] https://github.com/domkowald/LFM1b-analyses

[5] Predict the rating of all unobserved user-item combinations which can be seen as "true" ratings.
[6] Provide the top-N recommendations to each user based on all unobserved user-item interactions.
[7] Compute $GAP(g)_r$ (see Equation (3)) for each user group.
[8] Compute $\Delta GAP(g)$ (see Equation (4)) for each user group.
[9] Append the user-item combinations from the recommendations with their respective "true" rating to the dataset as new interactions.
[10] To simulate the feedback loop for $M$ iterations, steps [1] - [9] are repeated $M - 1$ times.

The above steps provide the evolution of $\Delta GAP$ (i.e., *dynamic-$\Delta GAP$*) over time for different user groups. This is demonstrated in section 4.3.2 below.

### 3.3 Between-group GAP

Currently, $\Delta GAP$ is formulated as the relative difference between $GAP_r$ (recommendations) and $GAP_p$ (user profiles). This results in $\Delta GAP$ ranging from $(-1, \infty)$, with negative values suggesting recommendations being less popular than user profiles, and vice versa for positive values (see section 2.3.2). Whereas, this approach allows for a comprehensive interpretation of within-group unfairness, it proves challenging to visualise unfairness between different groups. Therefore, we propose a revised formula of $\Delta GAP$ that can be used in *Between-group GAP* metric.

Since $GAP_r$ and $GAP_p$ are both average item popularity scores, under the assumption that items can only be interacted with once (i.e., users can only provide one rating to an item), we subtract 1 from their respective values to calculate the average *item non-popularity scores*, i.e.,

$$\Delta GAP_{revised} = \frac{1 - GAP_r}{1 - GAP_p}, \quad (6)$$

the values of $\Delta GAP_{revised}$ range from 0 to $\infty$. $\Delta GAP_{revised} = 1$, when $GAP_p = GAP_r$, implying the non-popularity score of items is the same in the recommendations as in the profiles of a user group. When $\Delta GAP_{revised} < 1$, the popularity of items in the recommendations are higher than the popularity of items in the user profiles. Whereas, when $\Delta GAP_{revised} > 1$, the recommended items are less popular than in the user profiles.

The benefit of this approach is that it allows taking into consideration the impact of popularity bias when comparing groups. To illustrate this, consider the following situations:
- **Situation 1:** Items recommended to a group are 50% more popular than items in their user profiles.
- **Situation 2:** Items recommended to a group are 50% less popular than items in their user profiles.

From the perspective of calibration fairness, both situations are similar, i.e., items recommended to a user group differs by 50% from their user profiles. However, based on the definition of popularity bias, where the over-recommendation of popular items leads to "the rich getting richer, the poor getting poorer", we argue that the situation 1 is more unfair compared to situation 2.

The formulation of $\Delta GAP_{revised}$ metric allows us to take this argument into account when comparing two sensitive groups. To compute the level of popularity bias between two groups we propose the *Between-group GAP* metric as follows:

$$BetweenGroup\ GAP(g, h)$$
$$= \left| \frac{\Delta GAP_{revised}(g) - \Delta GAP_{revised}(h)}{mean(\Delta GAP_{revised}(g), \Delta GAP_{revised}(h))} \right|, \quad (7)$$

where $g$ and $h$ are two sensitive user groups. *Between-group GAP* ranges from 0 to 2, with a perfect situation $\Delta GAP_{revised}(g) = \Delta GAP_{revised}(h)$ and *Between-group GAP* $= 0$. In other words, the level of unfairness towards group $g$ is the same as the level as the level of unfairness towards group $h$.

Since the metric aims to compare two sensitive group, the absolute difference is taken to avoid specifying one group as baseline. The output of *Between-group GAP* can be interpreted as the level of unfairness between two user groups. The higher the value of *Between-group GAP* is, the further apart the level of unfairness towards each group is. This is demonstrated in section 4.3.3 below.

### 3.3 Cosine Similarity

*Cosine similarity* is a measure that is widely used to compute the similarity between two vectors in a multi-dimensional space (Kirişci, 2023). In the context of popularity bias, *cosine similarity* can be used to calculate the similarity between frequency distributions of recommended items between two sensitive user groups. When deploying this metric in a dynamic setting, it may provide additional insights into the differential treatment in recommendations provided to two sensitive groups. Exploring the use of cosine similarity in addition to previously mentioned fairness metrics could enhance our understanding of popularity bias in RecSys.

The proposed *cosine similarity* metric to measure popularity bias in a dynamic setting is computed as following within each feedback iteration:

[1] For each sensitive group, generate a vector of zeros of length equal to the number of items in the dataset. For example, if a dataset consists of 5 items, the initial vector is [0, 0, 0, 0, 0].

[2] Update each element corresponding to an item in the vector with the number of times the item has been recommended to the sensitive group under consideration. For example, if item 2 is recommended 3 times, the updated vector would be [0, 3, 0, 0, 0].

[3] Normalise the vectors by the total number of users in the group, to take different group sizes into account.

[4] The two normalised vectors corresponding to two sensitive groups under consideration are then used to compute the *cosine similarity* of the frequency of recommended items between two sensitive groups.

The value of *cosine similarity* indicates the degree of similarity between the frequency of recommended items between two user groups. In an ideal situation, the *cosine similarity* equals 1 ; in the context of popularity bias this implies that the two user groups receive recommendations with similar frequencies.

## 4 DEMONSTRATIONS

Metrics that have been proposed in section 3 to assess popularity bias in dynamic setting and at user group level are demonstrated in this section.

### 4.1 Datasets

In academic research on RecSys, a variety of datasets have gained recognition as benchmark datasets such as the MovieLens 1M dataset, the Netflix dataset, the Amazon Product Datasets, and the Yelp dataset (Deldjoo et al., 2023; Lin et al., 2022; Singhal et al., 2017). As metrics of popularity bias for different user groups is in focus, datasets containing demographic features, such as gender, ethnicity, or education level has been used in this paper. Both the MovieLens and Yelp datasets possess sensitive features such as demographics and gender; therefore, these two datasets have been selected for demonstration.

The MovieLens dataset was developed by the GroupLens[5] team at the University of Minnesota, while the Yelp dataset was created by Yelp[6] having a subset of their businesses, reviews, and user data. In the pre-processing stage of the datasets, users and items with fewer than 10 interactions were removed to ensure an adequate level of data density and reliability (Lin et al., 2022).

### 4.2 Simulation Dynamic Setting

To simulate dynamic data, an approach similar to that of Sun et al. (2019) and Zhu et al. (2021) is adopted. First, a random sample of 1,000 users is taken from the original dataset, resulting in the dimensions presented in Table 1. In this table, 'items' represent the number of unique items present in the sample, 'ratings' the number of observed interactions (i.e., the observed user-item-rating combinations), and 'density' the indicator for the density of the user-item matrix calculated by dividing the number of observed interactions by the maximum number of possible interactions.

Table 1: Dimensions of selected datasets

| Dataset | Users | Items | Ratings | Density |
|---|---|---|---|---|
| MovieLens | 1,000 | 3,214 | 161,934 | 0.05 |
| Yelp! | 1,000 | 1,272 | 74,527 | 0.06 |

In the first feedback iteration, a specific RecSys algorithm is trained and used to predict the rating for all unobserved user-item combinations. These predictions are classified as the true rating for that combination to be used for the next iteration.

Then the top-10 recommendations are provided to each user in the dataset and user-item combinations of the recommendations is appended with their corresponding "true" ratings. In each subsequent iteration, the RecSys algorithm is trained using the appended dataset to predict the ratings for all unobserved interactions and then used to provide each user with their top-10 recommendations and the results are appended to the dataset. To study long term effects this process of simulating feedback iterations is repeated 40 times. Asymptotic behaviour is already observed at these iterations.

---

[5] https://grouplens.org/

[6] https://www.yelp.com/dataset/

Table 2: RecSys model details

| Dataset | Algorithm | Optimal Hyperparameters | | | | RMSE |
|---|---|---|---|---|---|---|
| MovieLens | SVD | Epochs: 50 | Factors: 150 | LR: 0.005 | RT: 0.05 | 0.90 |
| | NMF | Epochs: 100 | Factors: 150 | | | 0.89 |
| | userKNN | K: 20 | Metric: mean-squared deviation | | | 0.95 |
| | itemKNN | K: 75 | Metric: mean-squared deviation | | | 0.95 |
| Yelp! | SVD | Epochs: 10 | Factors: 75 | LR: 0.005 | RT: 0.05 | 0.92 |
| | NMF | Epochs: 100 | Factors: 150 | | | 0.94 |
| | userKNN | K: 50 | Metric: mean-squared deviation | | | 0.97 |
| | itemKNN | K: 50 | Metric: mean-squared deviation | | | 0.97 |

*Note: Epoch is a single pass through the dataset during training, factor the number of latent user and item factors used in the model, learning rate (LR) the hyperparameter determining the size of the steps taken during optimisation affecting how quickly the model converges, regularization term (RT) the penalty term added to the loss function to prevent overfitting, K the number of neighbours that are taken into account for aggregation, and metric the method for distance computation.*

### 4.3 Demonstration of Proposed Metrics of Popularity Bias

To compute the proposed metrics of popularity bias, multiple RecSys algorithms were trained to assess how these metrics vary per algorithm over time. Algorithms trained were Singular Value Decomposition (SVD) (Koren et al., 2009), Non-Negative Matrix Factorization (NMF) (Lee & Seung, 2000), user-based K-Nearest Neighbors (userKNN), and item-based K-Nearest Neighbors (itemKNN) (Adomavicius & Tuzhilin, 2005). Table 2 presents the hyperparameters of the trained RecSys models.

#### 4.3.1 Within-group-Gini coefficient

In this section, results of the metric *within-group-Gini coefficient* is presented for two sensitive user groups, males and females. Note, if a model consistently provides fewer or more diverse recommendations to a specific user group, it will lead to diverging values of this metric between the groups. Figure 1 presents the *within-group-Gini coefficient* of males and females for the MovieLens and Yelp datasets. The gap between the genders is visible more in MovieLens than in Yelp due to larger preferential differences between groups in movies than in their choice of restaurants. Additionally, 4 observations are made in Figure 1.

First, when evaluating differences between genders, consistent trends across both datasets are observed. Initially, in the first iteration, the metric's value for males is higher than for females. This disparity is more prominent in the MovieLens dataset (see Figure 1), but a similar trend also exists within the Yelp dataset, although with a smaller difference. These findings indicate that in the original dataset, males have interacted with a smaller diversity of items compared to females.

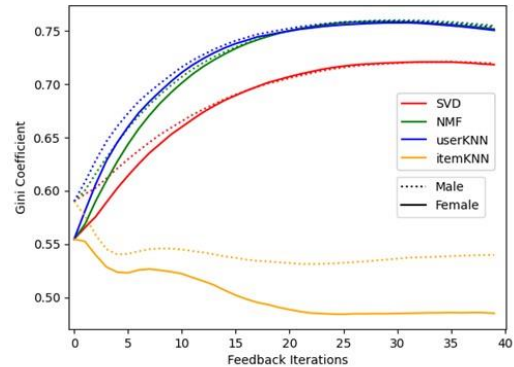

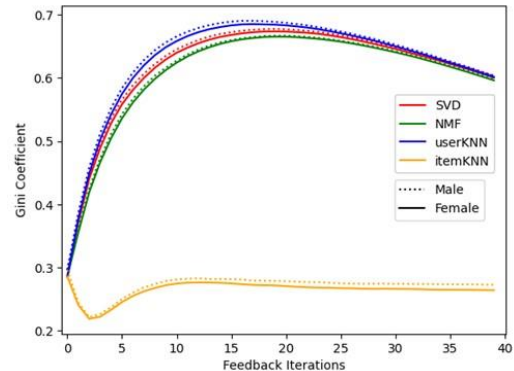

Figure 1: Results of within-group-Gini coefficient for male and female over a feedback loop per RecSys model

Second, in both datasets across both groups, except for in itemKNN, the Gini coefficient increases rapidly followed by an asymptote (MovieLens) or a slow decline (Yelp). This is expected because over time the propensity of recommending only popular items increases. Hence the diversity of recommendations decreases.

Third, the initial separation observed in values of *within-group-Gini coefficient* between the groups decreases over time, especially for MovieLens. This implies that over time, males and females are provided with increasingly similar recommendations in terms of item diversity. This does not suggest that the recommended items are the same between the two user groups, but the ascending order of distribution of item popularity scores are similar.

Fourth, itemKNN's behavioral difference compared to the other algorithms is expected as recommendations made in itemKNN are based on similarity in item features and not on user similarity. The difference between *within-group-Gini coefficient* increases over time between the groups, suggesting that over time, females are being recommended a more diverse set of items compared to males.

### 4.3.2 Dynamic-$\Delta GAP$

When applying *dynamic-$\Delta GAP$*, results (see Figure 2) reveal extreme values in the initial iterations. This is expected as user profiles are directly being compared with recommendations provided in the first iteration. It is also observed that initial values of *dynamic-$\Delta GAP$* are negative, suggesting that the initial group-recommendations were less popular than the average user profiles of the respective group. This turbulent starting phase is attributed to the "cold start" problem within RecSys (Volkovs et al., 2017), where users with relatively limited observed interactions in their user profiles exert a significant influence on the average popularity of their user group.

Except for itemKNN, the algorithms appear to converge to a negative *dynamic-$\Delta GAP$* value. This convergence to a negative value is because, SVD, NMF and userKNN are prone to over recommending popular items, which leads to an increase in the average item popularity in the user profiles for the initial iterations. After this initial phase these algorithms are left with a pool of less popular items to recommend. Hence, the average item popularity of the recommended items after the initial phase is lower than the average item popularity in the user profiles, leading to the convergence of *dynamic-$\Delta GAP$* to a negative value. On the contrary, itemKNN is based on recommending items with similar features.

Therefore, over time it recommends similar items to the user profiles. This leads to the average popularity of the recommended items to be similar to the average item popularity in the user profiles resulting in *dynamic-$\Delta GAP$* converging towards zero.

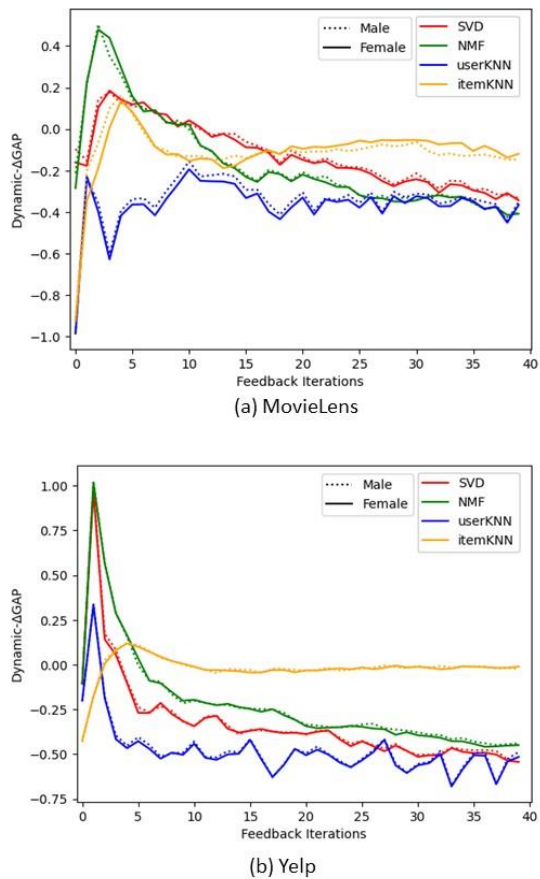

Figure 2: Results of dynamic-$\Delta$GAP for male and female over a feedback loop per RecSys model.

Lastly, with regards to the sensitive user groups, minor variations in *dynamic-$\Delta GAP$* is observed for MovieLens dataset, however, for Yelp dataset this variation is negligible. Similar to *within-group-Gini coefficient*, this difference in the two datasets is due to larger preferential differences between males and females in movies than in their choice of restaurants.

### 4.3.3 Between-group GAP

*Between-group GAP* is demonstrated using six hypothetical scenarios using $\Delta GAP_{revised}$ in a dynamic setting for male and female user groups. Thereby, it is validated that over-recommendation of popular items is more unfair than the over-recommendation of non-popular items.

Table 3: Values of $Between-group\ GAP$ for different scenarios of over-recommending popular or non-popular items for user groups $g$ and $h$.

|   | ItemPopularity(g) | ItemPopularity(h) | $\Delta GAP_{revised}(g)$ | $\Delta GAP_{revised}(h)$ | BetweenGroup GAP(g, h) |
|---|---|---|---|---|---|
| 1 | + 50% | + 50% | 0.67 | 0.66 | 0.00 |
| 2 | 0% | - 50% | 1.00 | 1.33 | 0.28 |
| 3 | 0% | + 50% | 1.00 | 0.67 | 0.40 |
| 4 | - 20% | + 10% | 1.13 | 0.93 | 0.19 |
| 5 | - 10% | + 20% | 0.07 | 0.87 | 0.21 |
| 6 | - 50% | + 50% | 1.33 | 0.67 | 0.67 |

To illustrate this, six hypothetical scenarios are presented in Table 3. In each scenario we first define the difference of the item popularity within recommendations compared to item popularity of user profiles. To illustrate, an "ItemPopularity" of +50% represents recommendations being 50% more popular than their user profiles (e.g., $GAP_p = 0.4$ results in $GAP_r = 0.6$); an "ItemPopularity" of -50% represents recommendations being 50% less popular than their user profiles.

Scenario 1 represents a perfect situation where both groups $g$ and $h$ are treated equally with both receiving recommendations that are 50% more popular than their profiles. Scenarios 2 and 3 demonstrate how over-recommending popular items is considered more unfair than over-recommending unpopular items. In scenario 2, group $g$ receives a perfect recommendation and group $h$ is recommended items that are 50% less popular than their user profiles, resulting in a $BetweenGroup\ GAP$ of 0.2857. In scenario 3, group $g$ also receives a perfect recommendation, but group h is recommended items that are 50% more popular than their user profiles instead, resulting in a $BetweenGroup\ GAP$ of 0.4. Even though in both scenarios one group receives recommendations that are 50% off their user profiles, it illustrates that over-recommending popular items is considered more unfair. Scenarios 4 and 5 illustrate the same but with both groups receiving unfair recommendations and for smaller differences between groups.

The results of $BetweenGroup\ GAP$ for the MovieLens and Yelp datasets are presented in Figures 3a and 3b, respectively. For the Yelp dataset, consistent and relatively low values of the metric is observed over time, indicating that all the four models exhibit relatively low degree of unfairness towards both males and females for the Yelp dataset. However, the MovieLens dataset shows an increasing trend in between-group unfairness, in particular for userKNN and NMF models (see Figure 3a). These models exhibit a distinct and noticeable pattern of growing disparities in treatment between males and females, resulting in a $BetweenGroup\ GAP$ value of 0.1. Referring to Table 3, a value of 0.1 indicates group $g$ receives perfect recommendations when compared to their profiles (0% item popularity), while group $h$ receives recommendations that are 14.3% more popular than their profiles.

These findings highlight the metric's potential in capturing and highlighting the emergence of unequal treatment between sensitive user groups. Furthermore, the metric demonstrates ability to differentiate between the fairness-performance of the four algorithms used in this study.

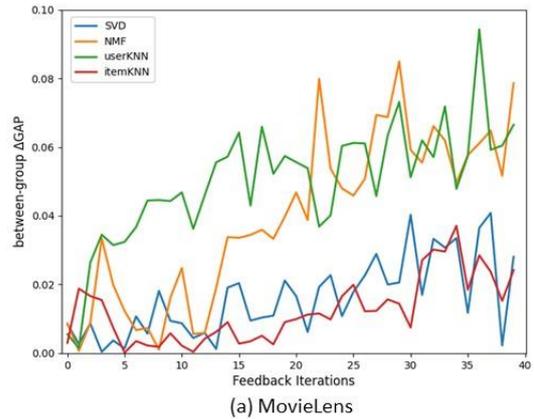

(a) MovieLens

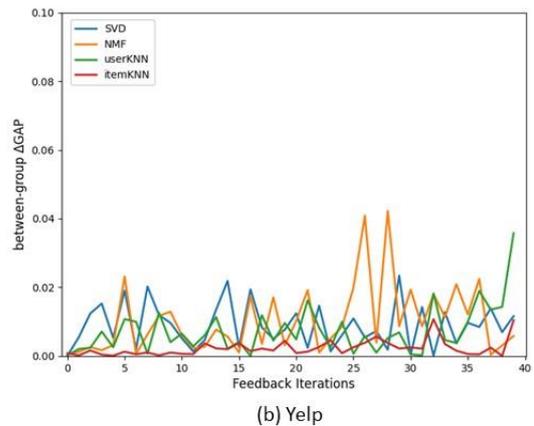

(b) Yelp

Figure 3: Results of between-group GAP for male and female over a feedback loop per RecSys model.

### 4.3.4 Cosine Similarity

The proposed metric for popularity bias, *cosine similarity,* measures the degree of similarity between the frequency of recommended items among two user groups. Values close to 1 represent a RecSys model recommending items with similar frequencies between user groups. Whereas values towards 0 represent the model recommending different items between groups.

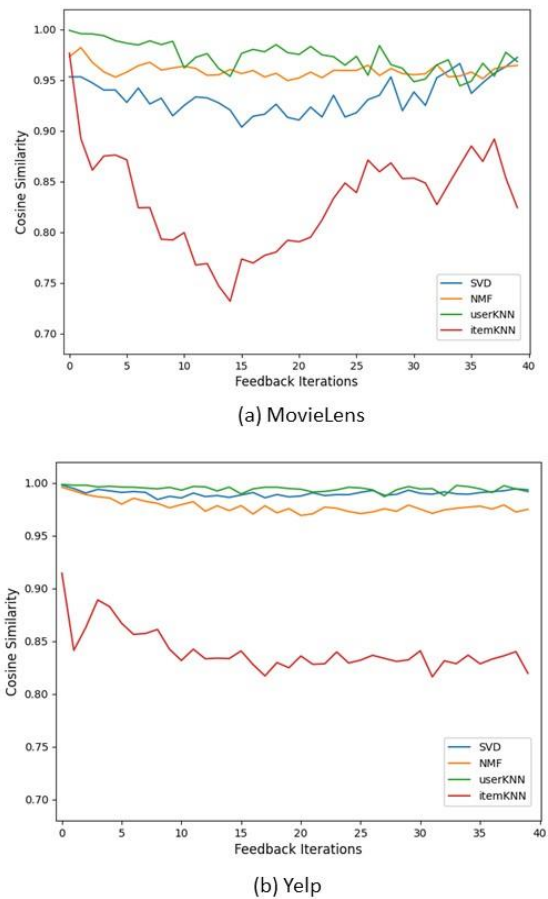

Figure 4: Results of cosine similarity for male and female over a feedback loop per RecSys model.

Figure 4 presents the results for the MovieLens and Yelp datasets. It is observed that the SVD, NMF, and userKNN models have cosine similarity close to 1, indicating a high and stable degree of similarity in recommendations between males and females over time. Conversely, the results for itemKNN suggest that the similarity of recommendations between males and females decreases over time. The results of itemKNN illustrate the importance of evaluating RecSys in a dynamic setting, as the similarity in the first iteration (i.e., a static setting) is significantly higher than in later iterations. The results presented demonstrate that the Yelp dataset presents more stable results than the MovieLens dataset, providing insights into the varying impact of different datasets on the performance of RecSys models.

It is important to note that this *Group-cosine similarity* has limitations due to potential differences in preferences between groups, rendering it insufficient as a standalone fairness metric. Consequently, future research should consider comparing the sorted normalised item popularity distributions between groups. This will enable evaluation of item popularity distributions among groups and can potentially reveal if one user group is presented with a more diverse set of items, offering valuable insights into the extent of recommendation diversity.

## 5 DISCUSSION & CONCLUSION

Popularity bias in RecSys leads to inequality in treatment between users or user groups due to over-recommendation of popular items. This bias arises from the disproportionate favouring of popular items leading to limited recommendation diversity and the potential exclusion of relevant but less popular items to certain users or groups. Therefore, it is important to quantify and track such biases in RecSys.

The commonly used metrics to measure popularity bias in RecSys are Gini coefficient, Delta group average popularity ($\Delta GAP$), and Generalized entropy index ($GEI$). Gini coefficient has been deployed in more real-life dynamic setting which quantifies inequality in the distribution of item popularity scores over time (Chong & Abeliuk, 2019; Zhu et al., 2021). However, this metric has only been used at a global level thus undermining the emergence of popularity bias between sensitive user groups. $\Delta GAP$, which measures the difference between the average item popularity in user group-recommendations and the average item popularity in user group-profiles has only been used in literature in static setting (Abdollahpouri et al., 2019). Biases in RecSys generally creeps in over-time asymmetrically in user groups, therefore it is important to extend the application of $\Delta GAP$ not only to dynamic setting but also to the context of different sensitive user groups. $GEI$ has been used to measure popularity bias both at group levels and in dynamic setting, however, Gini coefficient and $\Delta GAP$ are more commonly used due to their interpretability.

To consider the time evolution of popularity bias and its asymmetrical effects on different user groups,

four new metrics have been proposed, namely, *Within-group-Gini coefficient*, *Dynamic-$\Delta GAP$*, *Between-group GAP*, and *Group-cosine similarity*. *Within-group-Gini coefficient* evaluates the equality in the distribution of item popularities thus measuring how diverse recommendations are over time between sensitive user groups. Additionally, a new methodology to compute $\Delta GAP$, *Dynamic-$\Delta GAP$*, has been proposed where recommendations provided to the user are based on unobserved interactions in contrast to the original proposal of Abdollahpouri et al. (2019) and Kowald et al. (2020), in which users are recommended items from the testing data which they have already interacted with. The metric *BetweenGroup GAP* measures popularity bias resulting from the over-recommendation of popular or non-popular items. The metric *Group-cosine similarity* aims to assess the frequency of item recommendations among different user groups, specifically examining whether both groups are recommended the same items in equal proportions and can provide additional insights into the level of between-group popularity bias in Recsys.

The proposed metrics have been demonstrated using two commonly used datasets in academic research on RecSys, namely the MovieLens 1M dataset and the Yelp dataset with males and females as sensitive user groups. It is worthwhile to note that for a comprehensive understanding of time-evolution of popularity bias for different sensitive user groups in RecSys, it is advisable to use a combination of the metrics proposed in this paper. For example, as demonstrated in section 4, *BetweenGroup GAP* metric highlighted a growing disparity in treatment between males and females in the MovieLens dataset, whereas, *Within-group-Gini coefficient* metric revealed distinct trends in recommendation diversity among different RecSys models. It is also observed that the metric *BetweenGroup GAP* demonstrates ability to differentiate between the fairness-performance of the four algorithms used in this study.

## 6 FUTURE WORK

Future work involves implementing and evaluating the proposed metrics of popularity bias for more advanced deep-learning based RecSys systems to capture the complexity of industry-used models. Furthermore, additional approaches of the proposed metrics will be explored for their robust application, such as comparing sorted normalised item popularity distributions between different user groups to compute cosine similarity and finding an optimal method for incorporating $GAP_p$ in the calculation of *Dynamic-$\Delta GAP$* as discussed in section 5.